\documentclass[twocolumn]{aastex631}

\newcommand\submitms{y} 
\if\submitms y
\else
\usepackage{apjfonts}
\fi

\usepackage{ifthen}

\newcounter{fignum}

\newcommand{\comment}[1]{}
\newcommand{\ttt}[1]{$\times 10^{#1}$}

\usepackage{amsmath}
\usepackage{natbib}
\bibpunct[, ]{(}{)}{,}{a}{}{,}
\usepackage{appendix}
\usepackage{gensymb}
\usepackage{breqn}

\providecommand{\adsurl}[1]{\href{#1}{ADS}}

\DeclareSymbolFont{UPM}{U}{eur}{m}{n}
\DeclareMathSymbol{\umu}{0}{UPM}{"16}
\let\oldumu=\umu
\renewcommand\umu{\ifmmode\oldumu\else$\oldumu$\fi}
\newcommand\micro{\umu}

\newcommand\microns{\micro m}

\shorttitle{Eclipses of WASP-34b}
\shortauthors{Challener {\em et al.}}

\begin{document}

\title{Spitzer Dayside Emission of WASP-34b}
\author[0000-0002-8211-6538]{Ryan C. Challener}
\affiliation{Planetary Sciences Group, Department of Physics, 
  University of Central Florida, Orlando, FL 32816-2385}
\affiliation{Department of Astronomy, University of Michigan,
  1085 S. University Ave., Ann Arbor, MI 48109, USA}

\author[0000-0002-8955-8531]{Joseph Harrington}
\affiliation{Planetary Sciences Group, Department of Physics, 
  University of Central Florida, Orlando, FL 32816-2385}

\author[0000-0002-1347-2600]{Patricio E. Cubillos}
\affiliation{Planetary Sciences Group, Department of Physics, 
  University of Central Florida, Orlando, FL 32816-2385}
\affiliation{Space Research Institute, Austrian Academy of Sciences,
  Graz, Austria}

\author[0000-0002-0769-9614]{Jasmina Blecic}
\affiliation{Planetary Sciences Group, Department of Physics, 
  University of Central Florida, Orlando, FL 32816-2385}
\affiliation{New York University Abu Dhabi, Abu Dhabi, United Arab Emirates}

\author[0000-0002-3456-087X]{Barry Smalley}
\affiliation{Astrophysics Group, Keele University, Staffordshire ST5 5BG, UK}


\email{rchallen@umich.edu}

\begin{abstract}
  We analyzed two eclipse observations of the low-density transiting,
  likely grazing, exoplanet WASP-34b with the \textit{Spitzer Space
    Telescope's} InfraRed Array Camera (IRAC) using two techniques to
  correct for intrapixel sensitivity variation: Pixel-Level
  Decorrelation (PLD) and BiLinearly Interpolated Subpixel Sensitivity
  (BLISS). When jointly fitting both light curves, timing results are
  consistent within 0.7$\sigma$ between the two models and eclipse
  depths are consistent within 1.1$\sigma$, where the difference is
  due to photometry methods, not the models themselves. By combining
  published radial velocity data, amateur and professional transit
  observations, and our eclipse timings, we improved upon measurements
  of orbital parameters and found an eccentricity consistent with zero
  (0.0). Atmospheric retrieval, using our Bayesian Atmospheric
  Radiative Transfer code (BART), shows that the planetary spectrum
  most resembles a blackbody, with no constraint on molecular
  abundances or vertical temperature variation. WASP-34b is redder
  than other warm Jupiters with a similar temperature, hinting at
  unique chemistry, although further observations are necessary to
  confirm this.

\end{abstract}


\section{INTRODUCTION}
\label{introduction}

Relative system flux variations, during planetary and stellar
occultations, are the primary way we characterize exoplanetary
atmospheres.  Eclipse observations, when the planet passes behind the
star, reveal temperature and atmospheric composition of the planet's
dayside, and eclipse ephemerides constrain planetary orbital
eccentricity.

In this work, we analyzed two \textit{Spitzer Space Telescope}
\citep{WernerEtal2004apjsSpitzer} InfraRed Array Camera
\citep[IRAC][]{FazioEtal2004apjsIRAC} eclipse observations of the
exoplanet WASP-34b. WASP-34b is a hot Jupiter on a potentially-grazing
orbit around a Sun-like star. Its mass of $0.57 \pm 0.03$ Jupiter
masses \citep{KnutsonEtal2014apjRVsearch} and radius of $1.22 \pm
0.08$ Jupiter radii \citep{SmalleyEtal2011aapWASP34b} imply a very low
density of $\sim 0.43 \pm 0.01$ g/cm$^3$. This places WASP-34b in the
top 0.8\% least dense planets with a measured mass and radius, per the
NASA Exoplanet Archive (exoplanetarchive.ipac.caltech.edu).

IRAC exhibits several systematic effects which must be carefully
removed. Of particular interest for this work, there is a correlation
between target position and flux due to subpixel gain variation in the
detector. Several methods have been used to deal with this effect,
including polynomial maps
(e.g. \citealp{CharbonneauEtal2005apjTrES1}), BiLinearly Interpolated
Subpixel Sensitivity maps (BLISS,
\citealp{StevensonEtal2012apjBLISShd149b}), Pixel-Level Decorrelation
(PLD, \citealp{DemingEtal2015apjHATP20pld}), Independent Componenet
Analysis (ICA, \citealp{MorelloEtal2015apjICAgj436}), and Gaussian
Processes (GP, \citealp{GibsonEtal2012mnrasGP}). We measure eclipse
depths and timings utilizing both BLISS and PLD, which have been shown
to be among the most accurate methods
\citep{IngallsEtal2016ajSpitzerSystematics}.


This paper is organized as follows: in Section \ref{sec:obs} we
present the observations, in Section \ref{sec:analysis} we describe
our data analysis procedure, in Section \ref{sec:joint} we
  discuss a simultaneous fit to both light curves, in Section
\ref{sec:orbit} we fit orbital models to our light curve results, in
Section \ref{sec:atm} we present atmospheric retrievals based on
measured eclipse depths, in Section \ref{sec:discussion} we
  discuss WASP-34b in the context of other similar planets, and in
Section \ref{sec:conclusions} we lay out our conclusions.

\section{OBSERVATIONS}
\label{sec:obs}

We observed WASP-34 once with each of the 3.6 and 4.5
\microns\ photometric filters available during the warm
\textit{Spitzer} mission, as part of program 60003 (PI: Harrington).
Each observation spanned $\sim 7$ hours, such that the WASP-34b
eclipses would occur roughly in the middle and there would be enough
baseline to characterize and remove the \textit{Spitzer} systematic
effects. The two observations occurred 8 days apart, on July 19 and
July 27 2010, or two orbits of WASP-34b. We used the 0.4 second
exposure time for both observations.



\section{DATA ANALYSIS}
\label{sec:analysis}

The challenge with \textit{Spitzer} observations lies in correcting
the telescope's systematic effects. The InfraRed Array Camera (IRAC,
\citealp{FazioEtal2004apjsIRAC}) was designed for 1\% relative flux
precision, but exoplanet eclipse observations are of order 0.1\%.  We
are able to achieve $\sim$0.01\% precision with a careful treatment of
correlated noise using our Photometry for Orbits, Eclipses, and
Transits code (POET, \citealp{NymeyerEtal2011apjWASP18b,
  StevensonEtal2012apjBLISShd149b, BlecicEtal2013apjWASP14b,
  CubillosEtal2013ApjWASP8b, BlecicEtal2014apjWASP43b,
  CubillosEtal2014apjTrES1b, HardyEtal2017apjHATP13b,
  ChallenerEtal2021psjSystematics}).

POET applies a multitude of centroiding and photometry methods to
produce light curves. We use center-of-light, Gaussian, and
least-asymmetry \citep{LustEtal2014paspLeastAsym} centering
techniques. For photometry, we use three types of apertures: fixed,
where the size of the aperture does not change over the course of an
observation; variable, where the size of the aperture is adjusted for
changes in the width of the point-spread function (PSF) according to
the ``noise pixels'' \citep{LewisEtal2013apjHATP2b}; and elliptical,
where we use an elliptical aperture with $x$ and $y$ widths dependent
on a Gaussian fit to the star in every frame
\citep{ChallenerEtal2021psjSystematics}. We try fixed aperture radii
from 1.5 -- 4.0 pixels in 0.25 pixel increments. For variable
apertures, we use radii described by

\begin{equation}
  \label{eqn:varrad}
  R_{var} = a\sqrt{N}+b,
\end{equation}

\noindent
where $N$ is the noise pixel measurement for a given frame, $a$ ranges
from 0.5 -- 1.5 in 0.25 increments, and $b$ ranges from -1 -- 2 in
steps of 0.5. The elliptical apertures sizes are given by

\begin{align}
  R_x = a\sigma_x+b, \nonumber \\
  R_y = a\sigma_y+b, \label{eqn:ellrad}
\end{align}

\noindent
where $\sigma_x$ and $\sigma_y$ are the 1$\sigma$ widths of a Gaussian
fit to the star along the $x$ and $y$ axes, $a$ ranges from 3 -- 7 in
steps of 1, and $b$ covers -1 -- 2 in 0.5 increments.

POET chooses the best combination of centering and photometry methods
by minimizing the binned-$\sigma$ $\chi^2$ of the decorrelated
  photometry (hereafter $\chi^2_{\rm bin}$,
\citealp{DemingEtal2015apjHATP20pld}). When dominated by white noise,
the model standard deviation of normalized residuals (SDNR) should
reduce predictably with bin size as $1/\sqrt{\rm bin\ size}$. The
$\chi^2_{\rm bin}$ measures how well a line of slope $-1/2$ fits to
log(SDNR) vs.\ log(bin size), with a lower $\chi^2_{\rm bin}$
indicating less correlated noise. The optimal centering and photometry
methods are listed in Table \ref{tbl:centphot}.

\begin{deluxetable}{llll}
\tablecaption{\label{tbl:centphot} 
Centering and Photometry Parameters}
\tablehead{\colhead{Wavelength} & \colhead{Centering} & \colhead{Phot.} & \colhead{Ap. Rad.\tablenotemark{a}} \\
  \colhead{(\microns)} & \colhead{Method} & \colhead{Method} & \colhead{(pixels)}}
\startdata
BLISS\\
\tableline
3.6       & Gaussian          & Elliptical & 3.0+0.5  \\
4.5       & Least Asymmetry   & Fixed      & 2.5  \\
\tableline
PLD\\
\tableline
3.6       & Center-of-light   & Fixed      & 2.00  \\
4.5       & Gaussian          & Elliptical & 4.00+0.5  \\
\enddata
\tablenotetext{a}{Variable and elliptical aperture radii are given as $a+b$ (Equations \ref{eqn:varrad} and \ref{eqn:ellrad})}
\end{deluxetable}

There are two main systematics in IRAC photometry: a non-flat baseline
(``ramp'') and a position-dependent gain variation across the detector
at the subpixel level. The first can generally be corrected with a
linear or quadratic function, or occasionally no correction is
necessary.  To remove the position-dependent effect, we use both BLISS
\citep{StevensonEtal2012apjBLISShd149b} and PLD
\citep{DemingEtal2015apjHATP20pld},
separately. \cite{IngallsEtal2016ajSpitzerSystematics} compared seven
correlated-noise removal techniques and found these two methods to be
among the most accurate and reliable.

BLISS grids the detector into subpixels. We use the root mean square
(RMS) of the point-to-point variation in the $x$ and $y$ positions of
the target on the detector as the grid size in each respective
dimension.  BLISS then directly computes the detector gain variation
for each grid bin by assuming any remaining unmodeled effects are due
to gain variation. This is dependent on the centering method, as each
frame is assigned to a grid bin, and thus to a correction factor,
based on the position of the target.  With BLISS, the light curve
model is

\begin{equation}
  F(x,y,t) = F_s E(t) R(t) M(x,y)
\end{equation}

\noindent
where $F_s$ is the total system flux, $E(t)$ is an eclipse model,
$R(t)$ is a ``ramp'' model, and $M(x,y)$ is the BLISS map.

PLD notes that the motion of the target is encoded in the brightness
of the pixels; if the target moves left, pixels on the left brighten
and pixels on the right dim. It models the light curve as the sum of
several of the brightest pixels, multiplied by a weighting factor. The
pixel values are normalized at each frame such that their sum is one,
so that any time-dependent astrophysical effects are removed. We
choose to use the nine brightest pixels in this work. The light curve
model is then

\begin{equation}
  F(t) = F_s \left(\sum_{n=1}^{9} c_i \hat P_i^t  + R(t) + E(t) \right),
\end{equation}

\noindent
where $c_i$ are the pixel weights, $\hat P_i^t$ are the normalized pixel
values at time $t$, $R(t)$ is a ramp model, and $E(t)$ is an eclipse model.
PLD also bins the data in time and chooses the best binning level
using $\chi^2_{bin}$.

In this work, we try the following ``ramp'' functions with BLISS:

\begin{align}
  R(t) &= 1,\\
  R(t) &=                  r_1 (t - 0.5) + 1,\\
  R(t) &= r_2 (t - 0.5)^2 + r_1 (t - 0.5) + 1,
\end{align}

\noindent
where $r_i$ are free parameters and $t$ is in units of orbital phase,
where transit occurs at 0 orbital phase. With PLD, we instead use the
following functions, because PLD treats variations additively and
thus, the functions must be relative to 0:

\begin{align}
  R(t) &= 0,\\
  R(t) &=                  r_1 (t - 0.5),\\
  R(t) &= r_2 (t - 0.5)^2 + r_1 (t - 0.5).
\end{align}

\noindent
For the final fit, we choose the ramp model which results in the
lowest Bayesian Information Criterion (BIC, \citealp{Raftery1995BIC}),
given by

\begin{equation}
  \label{eqn:bic}
  \textrm{BIC} = \chi^2 + k\ \textrm{ln}\ n,
\end{equation}

\noindent
where $k$ is the number of free parameters and $n$ is the number of
data points. The BIC is a measure of goodness of fit with a penalty
for added free parameters. Relative model confidence is assessed as

\begin{equation}
  \label{eqn:biccomp}
  P_{21} = \textrm{exp}\left(-\frac{\textrm{BIC}_2 - \textrm{BIC}_1}{2}\right),
\end{equation}

\noindent
where model 2 has a larger BIC than model 1. Note that since the
BIC is dependent on the size of the data set, data binning
must be kept constant when comparing the BICs of different
models.

For the eclipse model $E(t)$ we use a version of the uniform source
model from \citet{MandelAgol2002apjlLightCurve}. Since WASP-34b is
potentially a grazing planet \citep{SmalleyEtal2011aapWASP34b}, we
account for a nonzero impact parameter, and thus fit to the maximum
depth of the eclipse (if it was not grazing), rather than the depth of
the feature in the light curve. Such a model is necessary to get an
accurate temperature measurement of the dayside of the planet. For a
planet smaller than its star, \cite{MandelAgol2002apjlLightCurve}
define the ratio of obscured light during a transit as $F^e(p,z) = 1 -
\lambda^e(p,z)$, where

\begin{equation}
  \label{eqn:ecl}
  \lambda^e = 
  \begin{cases}
    \frac{1}{\pi} \left(k_0p^2 + k_1 \right. \\ \quad \left. -\sqrt{\frac{4z^2 - (1+z^2-p^2)^2}{4}}\right), & 1 - p < z < 1 + p\\
    p^2, &           z \leq 1 - p\\
    0,       & \textrm{otherwise}    
  \end{cases}
\end{equation}

\noindent
where $k_0$ and $k_1$ are defined as

\begin{align}
  \label{eqn:k1k2}
  k_0 &= \arccos{\left(\frac{  p^2+z^2-1}{2pz}\right)},\\
  k_1 &= \arccos{\left(\frac{1-p^2+z^2  }{2z}\right)},
\end{align}

\noindent
$p^2$ is the area ratio of the planetary disk to the stellar disk
$R_p/R_s$, and $z$ is the distance, in stellar radii, from center of
the stellar disk to the center of the planetary disk, if both are
projected onto a plane perpendicular to the line of sight.

For eclipses, we rewrite this function to separate the depth of the
transit from the conditions of the piecewise definition. We note
that the area of overlap between the planetary and stellar disks
is

\begin{equation}
  A_{\rm over} = A_s \lambda^e(p,z),
\end{equation}

\noindent
where $A_s = \pi R_s^2$ is the area of the stellar disk. Then, the
area ratio of the obscured portion of the planetary disk to the total
planetary disk is

\begin{equation}
  A_{\rm rat} = \frac{\lambda^e(p,z)}{p^2}.
\end{equation}

\noindent
Then, if we define $D$ as the flux ratio of the planet to the star,
the eclipse function is

\begin{equation}
  E(t) = 1 - D\frac{\lambda^e(p,z)}{p^2}.
\end{equation}

\noindent
We compute $z$ as a function of time, eclipse midpoint, and impact
parameter, where we assume the planet moves at a constant velocity
behind the stellar disk dependent on the orbital period and semimajor
axis. The full eclipse model has parameters for eclipse midpoint,
planet-to-star flux ratio (maximum eclipse depth if non-grazing),
impact parameter $b$, orbital period $P$, stellar radius $R_s$,
planetary radius $R_p$, and orbital semi-major axis $a$.


In both observations the eclipse signals are too weak to constrain all
model parameters so we use Gaussian priors of $P = 4.3176782 \pm
0.0000045$ days, $b = 0.904^{+0.017}_{-0.014}$, $R_s = 0.93 \pm 0.12
R_{\odot}$, $R_p = 1.22^{+0.11}_{-0.08} R_{\rm J}$, and $a = 0.0524
\pm 0.0004$ AU \citep{SmalleyEtal2011aapWASP34b}. While this $b$ was
measured during transit, the planet's orbit is circular or nearly
circular \citep{KnutsonEtal2014apjRVsearch, BonomoEtal2017aandaGAPS},
so this is a reasonable assumption. Eclipse midpoint, planet-to-star
flux ratio, ramp parameters, and pixel weights when using PLD are left
free to vary with large parameter ranges and uninformative, uniform
priors.

We determined best fits using least-squares, and calculated
uncertainties with Markov-chain Monte Carlo (MCMC) utilizing
Multi-Core Markov-Chain Monte Carlo (MC$^3$,
\citealp{CubillosEtal2017ajRedNoise}). We rescale the data
uncertainties such that our fits have a reduced $\chi^2$ of 1, except
when comparing BICs of ramp models, as the rescaling forces a ``good''
fit when there may be none. We ran our MCMC until the chains satisfied
the Gelman-Rubin convergence test within 1\%
\citep{GelmanRubin1992stascGRTest}. We use the MCMC posterior
distribution of eclipse depths as a Monte Carlo sample to determine a
band-integral brightness temperature for each observation.

\subsection{\texorpdfstring{3.6 \microns}{3.6 um}}
\label{sec:wa034bs11}

\begin{figure*}[!ht]
\if\submitms y
   \setcounter{fignum}{\value{figure}}
   \addtocounter{fignum}{1}
   \newcommand\fignam{f\arabic{fignum}.pdf}
\else
   \newcommand\fignam{figs/bliss-lcs.pdf}
\fi
  
  \includegraphics[width=7in]{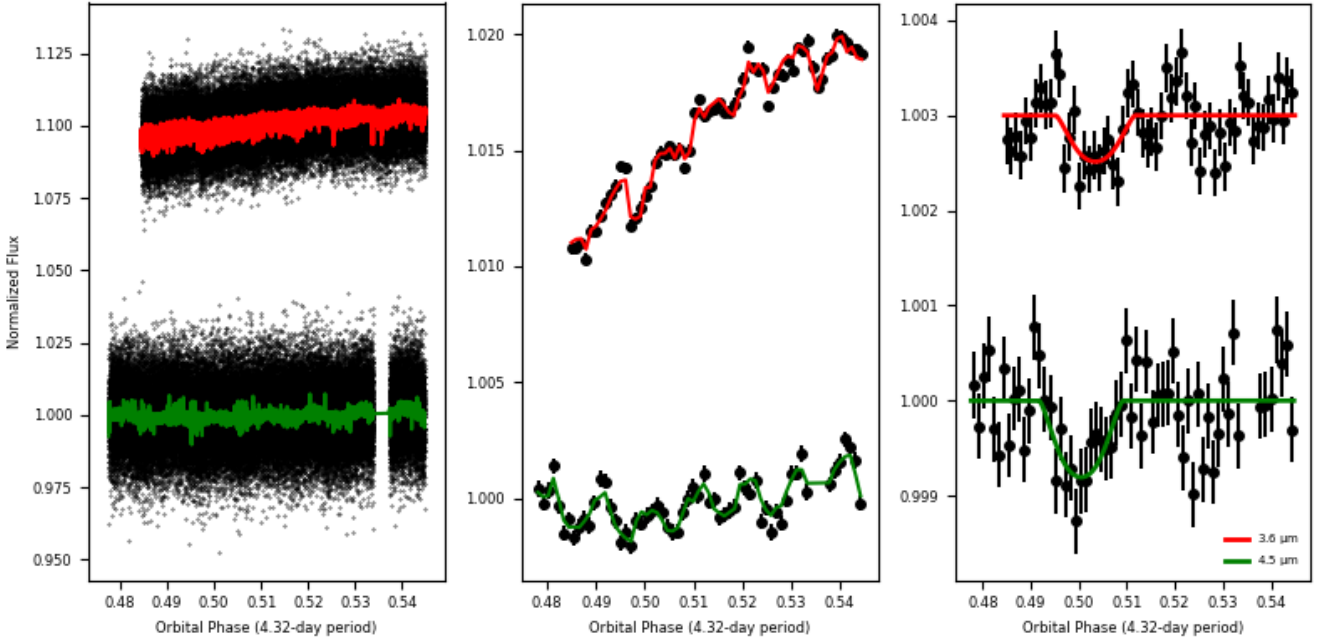}
  \caption{Individually-fit BLISS light curves of WASP-34. The 3.6
    \microns\ observation has been vertically offset for visual
    clarity. Note that we clipped out frames 49000 -- 52000 due to
    erratic sky levels and frames 53740 -- 53790 due to a reaction
    wheel spike. \textbf{Left:} Normalized raw photometry with
    best-fit models overplotted. \textbf{Middle:} Normalized binned
    photometry and binned best-fit models. \textbf{Right:} Normalized
    binned photometry and best-fit models with systematics divided out
    to highlight the eclipses.}
  \label{fig:blisslcs}
\end{figure*}

\begin{figure*}[!ht]
\if\submitms y
   \setcounter{fignum}{\value{figure}}
   \addtocounter{fignum}{1}
   \newcommand\fignam{f\arabic{fignum}.pdf}
\else
   \newcommand\fignam{figs/pld-lcs.pdf}
\fi
   
  \includegraphics[width=7in]{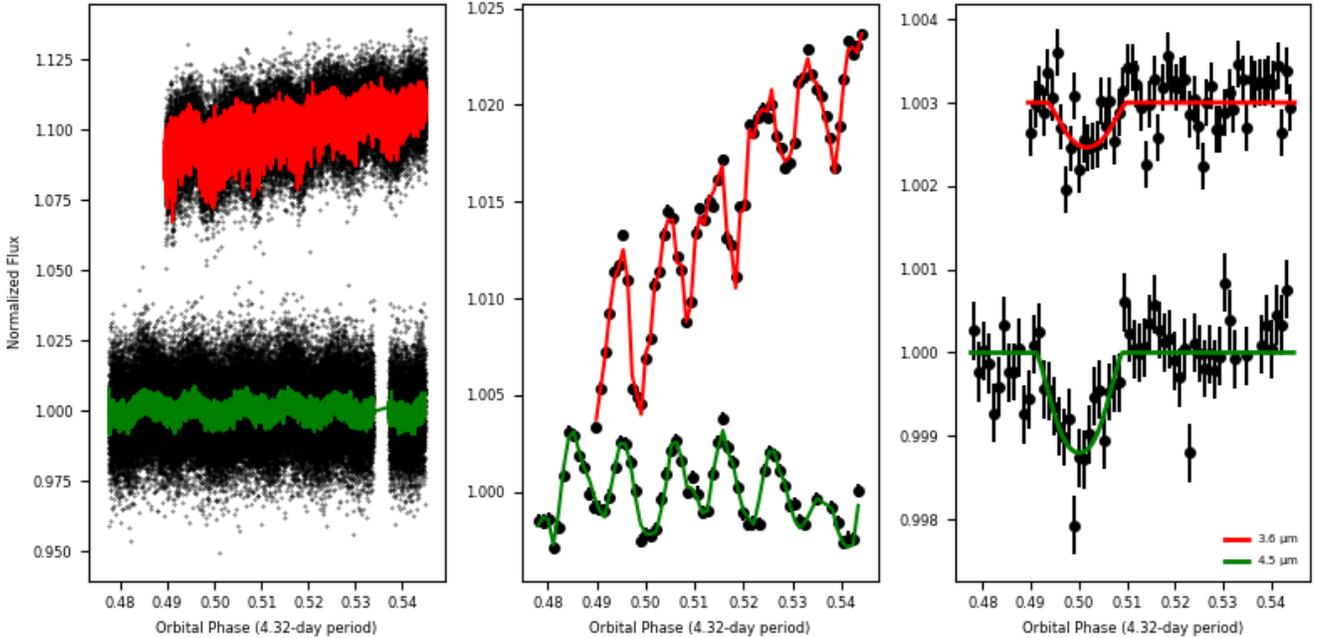}
  \caption{Individually-fit PLD light curves of WASP-34. The 3.6
    \microns\ observation has been vertically offset for visual
    clarity. Note that we clipped out frames 49000 -- 52000 due to
    erratic sky levels and frames 53740 -- 53790 due to a reaction
    wheel spike. \textbf{Left:} Normalized raw photometry with
    best-fit models overplotted. \textbf{Middle:} Normalized binned
    photometry and binned best-fit models. \textbf{Right:} Normalized
    binned photometry and best-fit models with systematics divided out
    to highlight the eclipses.}
  \label{fig:pldlcs}
\end{figure*}

Assuming a non-inclined orbit and a blackbody planet at its
zero-albedo, instantaneous heat redistribution equilibrium temperature
(1158 K), we expect a 3.6 \microns\ eclipse S/N of $<5$. Given that
WASP-34b's orbit is more likely grazing than not, and that systematic
effects are stronger at 3.6 \microns, it is unsurprising that this
detection is very weak. With BLISS, we determine an eclipse depth of
$560 \pm 154$ ppm centered at $2455396.68631 \pm 0.00345$
$\textrm{BJD}_{\textrm{TDB}}$. PLD finds an eclipse depth of $616 \pm
173$ ppm at $2455396.67882 \pm 0.00418$ $\textrm{BJD}_{\textrm{TDB}}$,
using a bin size of 8 frames. Figures \ref{fig:blisslcs} and
\ref{fig:pldlcs} show the BLISS and PLD fits. These depths correspond
to band-integrated brightness temperatures of $1257 \pm 109$
K and $1290 \pm 111$ K for BLISS and PLD, respectively. Table
\ref{tbl:ramps} lists the optimal ramp models for each
systematic-removal technique. We note telescope settling was
pronounced in this observation, so we clipped the first 10\% and
17.5\% of the data set for the BLISS and PLD fits, respectively.

\begin{deluxetable}{lrlrl}
\tablecaption{\label{tbl:ramps} 
  Ramp Model BICs}
\tablehead{ & \multicolumn{2}{c}{BLISS}               & \multicolumn{2}{c}{PLD} \\
\colhead{Ramp}     & \colhead{$\Delta$BIC} & \colhead{$P_{21}$}  & \colhead{$\Delta$BIC} & \colhead{$P_{21}$}}
\startdata
3.6 \microns\\
\tableline
None      &  546.1      & 2.61\ttt{-119}            & 317.5       & 1.14\ttt{-69}\\
Linear    &   99.0      & 3.18\ttt{-22}             &  14.6       & 6.76\ttt{-4}\\
Quadratic &    0.0      & ---                       &   0.0       & ---    \\
\tableline
4.5 \microns\\
\tableline
None      &    0.7      & 7.05\ttt{-1}              &   0.0       & --- \\
Linear    &    0.0      & ---                       &  10.1       & 6.41\ttt{-3} \\
Quadratic &   10.6      & 4.99\ttt{-3}              &  19.9       & 4.77\ttt{-5} \\
\enddata
\end{deluxetable}

\begin{figure}
\if\submitms y
   \setcounter{fignum}{\value{figure}}
   \addtocounter{fignum}{1}
   \newcommand\fignam{f\arabic{fignum}.pdf}
\else
   \newcommand\fignam{figs/pld-lcs.pdf}
\fi
   
\includegraphics[width=3.5in]{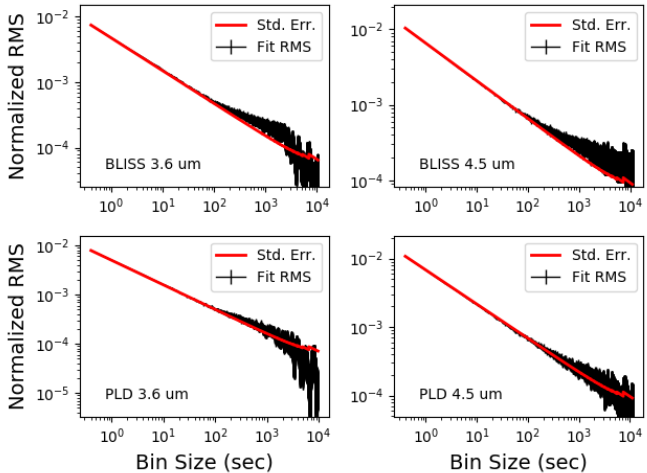}

\caption{Comparison of fit residual RMS vs.\ bin size with the
  expected standard error. If the fit residual RMS is above the
  standard error, there is correlated noise present at that time
  scale. For all four cases, the standard error is within the
  uncertainties of the residual fit RMS at nearly all bin sizes,
  indicating low residual correlated noise.}
\label{fig:rms}
  \end{figure}

The binned light curves show some potential residual correlated
noise. While our light-curve optimization methods minimize correlated
noise, we compare the residual RMS vs.\ different bin sizes with the
expected standard error in Figure \ref{fig:rms}. There is some
correlated noise present, but it is within 1$\sigma$ of the expected
standard error at nearly all bin sizes.

\subsection{\texorpdfstring{4.5 \microns}{4.5 um}}
\label{sec:wa034bs21}

Since the planet is brighter at 4.5 \microns\ than 3.6
\microns\ relative to the host star, here we expect a deeper
eclipse. Indeed, BLISS finds an eclipse depth of $895 \pm 201$ at
$2455405.30880 \pm 0.00327$ $\textrm{BJD}_{\textrm{TDB}}$, and PLD
finds an eclipse depth of $1312 \pm 147$ ppm at $2455405.30727 \pm
0.00253$ $\textrm{BJD}_{\textrm{TDB}}$, using a bin size of 16
frames. These depths correspond to band-integrated brightness
temperatures of $1279 \pm 103$ K and $1475 \pm 67$ K for BLISS and
PLD, respectively. Figures \ref{fig:blisslcs} and \ref{fig:pldlcs}
show the BLISS and PLD fits. Due to unusual sky level activity and a
reaction wheel spike, we removed frames 49000 -- 52000 and 53740 --
53790, respectively. Again, Table \ref{tbl:ramps} compares the ramp
models and Figure \ref{fig:rms} checks for residual correlated noise.

We note there is a $\sim$1.7$\sigma$ difference between these
eclipse depths. This is entirely due to differences in the selected
photometry methods. Regardless of PLD or BLISS, fixed aperture
photometry finds an eclipse depth of $\sim$850 ppm, whereas variable
and elliptical photometry produce an eclipse depth of $\sim$1300
ppm. Since the $\chi^2_{\textrm{bin}}$ prefers elliptical photometry
when using a PLD model, we present those results, but note that, at
least in this observation, the choice of photometry method impacts
results.

\section{Joint Light-Curve Modeling}
\label{sec:joint}

In an attempt to further constrain $b$, eclipse midpoint, and
planet-to-star flux ratio, we jointly fit to both light curves, with
both BLISS and PLD using the photometry listed in Table
\ref{tbl:centphot}. We use the same model parameterization scheme as
described above, but share $b$, $P$, $R_s$, $R_p$, $a$, and eclipse
midpoint (in orbital-phase space) between models of the 3.6
\microns\ and 4.5 \microns\ eclipses. With BLISS, we find $b = 0.907
\pm 0.016$, 3.6 \microns\ eclipse depth of $455 \pm 165$ pmm (1191
$\pm$ 129 K), 4.5 \microns\ eclipse depth of $868 \pm 196$ ppm (1261
$\pm$ 105 K), and an eclipse midpoint of $0.5018 \pm 0.0007$ orbital
phase.  Using the same configuration with PLD, we find $b = 0.907 \pm
0.015$, a 3.6 \microns\ eclipse depth of $606 \pm 147$ ppm (1299 $\pm$
98 K), a 4.5 \microns\ eclipse depth of $1283 \pm 310$ ppm (1463 $\pm$
127 K), and an eclipse midpoint of $0.5012 \pm 0.0006$ orbital
phase. The joint-fit eclipse midpoints are consistent with the
individual fits within $1.4\sigma$, and the eclipse depths are
consistent within $0.5\sigma$.

\section{Orbit}
\label{sec:orbit}

Eclipse observations, since they sample a different portion of the
orbit than transits, can significantly reduce uncertainties on
eccentricity, as well as detect eccentricity false positives in
radial-velocity (RV) data \citep{ArrasEtal2012mnrasRVtides}. We used
RadVel \citep{FultonEtal2018paspRadVel} to fit a Keplerian orbit to
the measured eclipse midpoint timings, published and amateur transit
ephemerides (var2.astro.cz/ETD/, Table \ref{tbl:tr}), and RV data
(Table \ref{tbl:rv}). None of the RV data occur during transit, so
there is no need to account for the Rossiter-McLaughlin effect.

\begin{deluxetable}{lllr}
\tablecaption{\label{tbl:tr} 
  WASP-34b Transit Observations}
\tablehead{\colhead{Time} & \colhead{Uncertainty} & \colhead{Reference\tablenotemark{a}}  \\ 
\colhead{(BJD$_{\textrm{TDB}}$)} & \colhead{(BJD$_{\textrm{TDB}}$)} & }
\startdata
2455739.92619     & 0.00117               & ETD: Curtis I.  \\
2455726.97299     & 0.0016                & ETD: Curtis I.  \\
2455631.97466     & 0.0013                & ETD: Evans P. \\
2455580.17290     & 0.00116               & ETD: Tan TG \\
2454647.55359     & 0.00064               & \cite{SmalleyEtal2011aapWASP34b}\\
\enddata
\tablenotetext{a}{ETD: Exoplanet Transit Database. We
  require that transits have a data quality of 3 or better.}
\end{deluxetable}

\begin{deluxetable}{llr}
  \tablecaption{\label{tbl:rv}
    WASP-34b Radial Velocity Data}
  \tablehead{\colhead{Time} & \colhead{RV} & \colhead{Reference\tablenotemark{a}} \\ 
    \colhead{(BJD$_{\textrm{TDB}}$)}  & \colhead{(m/s)} &}
\startdata
2455166.8246 &	49790.3 $\pm$	4.4 & 1\\
2455168.8191 &	49937.2 $\pm$	4.3 & 1\\
2455170.8439 &	49792.3 $\pm$	4.2 & 1\\
2455172.8246 &	49925.3 $\pm$	4.6 & 1\\
2455174.8495 &	49814.1 $\pm$	4.1 & 1\\
2455175.8487 &	49797.3 $\pm$	3.9 & 1\\
2455176.8235 &	49880.6 $\pm$	4.2 & 1\\
2455179.8425 &	49788.8 $\pm$	4.1 & 1\\
2455180.8566 &	49861.1 $\pm$	4.1 & 1\\
2455181.8219 &	49941.4 $\pm$	4.2 & 1\\
2455182.8521 &	49876.5 $\pm$	4.9 & 1\\
2455184.8554 &	49843.2 $\pm$	4.4 & 1\\
2455186.8299 &	49905.8 $\pm$	4.6 & 1\\
2455190.8509 &	49915.2 $\pm$	4.5 & 1\\
2455261.7740 &	49768.6 $\pm$	4.9 & 1\\
2455262.6724 &	49819.1 $\pm$	4.1 & 1\\
2455372.5078 &	49873.1 $\pm$	5.0 & 1\\
2455375.6020 &	49879.7 $\pm$	7.0 & 1\\
2455376.5170 &	49895.6 $\pm$	8.0 & 1\\
2455380.5170 &	49892.2 $\pm$	4.8 & 1\\
2455391.4971 &	49763.1 $\pm$	5.3 & 1\\
2455399.4719 &	49769.5 $\pm$	4.8 & 1\\
2455403.4683 &	49815.9 $\pm$	4.9 & 1\\
2455410.4719 &	49891.3 $\pm$	4.9 & 1\\
2455902.1333619 & -29.089 $\pm$ 1.561 & 2 \\
2455903.0804119 & 6.179 $\pm$ 1.576 & 2 \\
2455904.1423798 & -80.603 $\pm$ 1.542 & 2 \\
2455932.1205709 & -54.071 $\pm$ 1.664 & 2 \\
2456266.1084835 & 57.445 $\pm$ 1.547 & 2 \\
2456320.1059713 & -27.768 $\pm$ 2.146 & 2 \\
2456326.1395332 & 90.815 $\pm$ 2.212 & 2 \\
2456639.1639846 & 37.544 $\pm$ 1.684 & 2 \\
\enddata
\tablenotetext{a}{(1) \citealp{SmalleyEtal2011aapWASP34b} (2) \citealp{KnutsonEtal2014apjRVsearch}}
\end{deluxetable}

From long-term trends in the RV data, there is a candidate
  large-orbit companion in the system
  \citep{SmalleyEtal2011aapWASP34b, KnutsonEtal2014apjRVsearch}. We
  include this object to accurately model the RV data, although the
  new data in this work place no additional constraints on the
  companion. Our model includes terms for $e \textrm{cos} \omega$, $e
  \textrm{sin} \omega$, transit ephemeris $T_0$, orbital period, RV
  semi-amplitude $K$, RV zero point $\gamma$ (per instrument), and RV
  jitter $j$ (per instrument). Like \cite{KnutsonEtal2014apjRVsearch},
  we set $e \textrm{cos} \omega$ and $e \textrm{sin} \omega$ of the
  companion to 0.

We fit to both the BLISS and PLD results (individual fits from
Sections \ref{sec:wa034bs11} and \ref{sec:wa034bs21}, since using the
joint fits would force a conversion from orbital phase space to a
single Julian date, but the joint fit encompasses two eclipses) to
check for consistency (see Table \ref{tbl:orbit}). The two fits agree
well on all orbital parameters except $e_b \textrm{cos} \omega_b$,
which differs by 1.7$\sigma$. However, the uncertainty on the derived
$e$ is driven by the larger uncertainty on $e_b \sin \omega_b$, so
there is an insignificant difference in the recovered planetary
eccentricity. The addition of amateur transit timings and the eclipses
from this work improves the uncertainty on orbital period by 13\% over
\cite{KnutsonEtal2014apjRVsearch}.

\begin{figure}
\if\submitms y
   \setcounter{fignum}{\value{figure}}
   \addtocounter{fignum}{1}
   \newcommand\fignam{f\arabic{fignum}.pdf}
\else
   \newcommand\fignam{figs/ecchist.pdf}
\fi
   
  \includegraphics[width=3.25in]{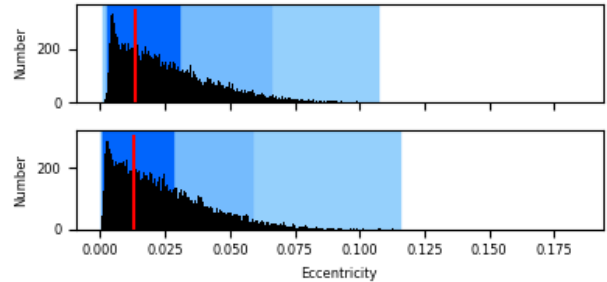}
  \label{fig:ecc}
  \caption{Eccentricity histograms derived from MCMC posterior
    distributions of $e\textrm{sin}\omega$ and
    $e\textrm{cos}\omega$. The red line marks the best-fit value, and
    the blue regions denote the 1, 2, and 3$\sigma$
    regions. \textbf{Top:} Posterior from the fit to the BLISS
    results. \textbf{Bottom:} Posterior from the fit to the PLD
    results.}
\end{figure}

The 1$\sigma$ uncertainty on $e$ indicates only a marginal detection
of eccentricity, consistent with \cite{KnutsonEtal2014apjRVsearch} and
\cite{BonomoEtal2017aandaGAPS}. However, the posterior distributions
show a 2--3$\sigma$ detection so we investigate the expected
circularization timescale for this planet and compare with the age of
the system. This timescale, from
\cite{GoldreichSoter1966IcarCircularization}, is given by

\begin{equation}
\label{eqn:tcirc}
\tau_e = \frac{4}{63} Q \left(\frac{a^3}{GM}\right)^{1/2}\left(\frac{m}{M}\right)\left(\frac{a}{R_{\textrm{p}}}\right)^5,
\end{equation}

\noindent
where Q is a tidal dissipation factor, typically $\sim10^6$ for hot
Jupiters \citep{Wu2005ApJValueofQ}, $a$ is orbital radius, $M$ is
stellar mass, $m$ is planetary mass, and $R_p$ is planetary radius.
Using $M = 1.01\ M_{\odot}$, $m = 0.59\ R_{\textrm{J}}$, and $a =
0.0524$ AU \citep{SmalleyEtal2011aapWASP34b}, we determine a
circularization timescale of $\sim$ 4\ttt{8}
years. \cite{SmalleyEtal2011aapWASP34b} note that lithium depletion in
WASP-34 indicates an age $\gtrsim$ 5 Gyr
\citep{SestitoRandich2005aapLiStellarAge}, implying that the planet's
orbit should have circularized. This is consistent with our results
within $\sim2\sigma$.

\begin{deluxetable*}{lccc}
\tablecaption{\label{tbl:orbit} 
WASP-34b Orbital Parameters}
\tablehead{&\colhead{BLISS}             & \colhead{PLD}               & \colhead{\cite{KnutsonEtal2014apjRVsearch}}}
\startdata
Fitted Parameters\\
\tableline
$e_b$ sin $\omega_b$                    & -0.013 $\pm$ 0.029          & -0.013  $\pm$ 0.028           & -0.001$^{+0.011}_{-0.017}$ \\
$e_b$ cos $\omega_b$                    &  0.0036 $\pm$ 0.0009        &  0.0016   $\pm$ 0.0008        & -0.0001$^{+0.0068}_{-0.0071}$ \\
$P_b$ (days)                            & 4.3176694 $\pm$ 0.0000038   & 4.3176694 $\pm$ 0.0000039     & 4.3176779 $\pm$ 0.0000045\\
$T_{0,b}$ (BJD$_{\textrm{TDB}}$)            & 2454647.55357 $\pm$ 0.00065 & 2454647.55357 $\pm$ 0.00065   & 2454647.55434$^{+0.00063}_{-0.00064}$\\
$K_b$ (m/s)                             & 71.0 $\pm$ 1.7              & 71.0 $\pm$ 1.7                & 71.1$^{+1.6}_{-1.7}$\\
$e_c$ sin $\omega_c$                    & 0                           & 0                             & 0 \\
$e_c$ cos $\omega_c$                    & 0                           & 0                             & 0 \\
$P_c$ (days)                            & 3990 $\pm$ 810              & 3960 $\pm$ 760               & 4093$^{+750}_{-520}$\\
$T_{0,c}$ (BJD$_{\textrm{TDB}}$)            & 2454612 $\pm$ 210           & 2454618 $\pm$ 200            & 2454589$^{+140}_{-190}$\\
$K_c$ (m/s)                             & 180 $\pm$ 60                & 179 $\pm$ 54                  & 189$^{+60}_{-35}$\\
$\gamma_{\rm CORALIE}$ (m/s)               & 50000 $\pm$ 62             & 49999 $\pm$ 57                & 141$^{+62}_{-37}$\\
$\gamma_{\rm HIRES}$ (m/s)                &  99 $\pm$ 62                & 97 $\pm$ 56                   & 108$^{+62}_{-37}$\\
$j_{\rm CORALIE}$ (m/s)                    & 6.1 $\pm$ 1.7              & 6.1 $\pm$ 1.7                 & ---\\
$j_{\rm HIRES}$ (m/s)                     & 1.7 $\pm$ 4.6               & 1.5 $\pm$ 4.2                 & ---\\
$j$ (m/s)                               & ---                        & ---                            & 3.2$^{+0.72}_{-0.6}$ \\
\tableline
Derived Parameters\\
\tableline
$e_b$                                   & 0.014$^{+0.017}_{-0.010}$     &  0.013$^{+0.015}_{-0.012}$        & 0.0109$^{+0.015}_{-0.0078}$\\
$\omega_b$ (\degree)                    & 286 $\pm$ 80               & 277 $\pm$ 86                  & 215$^{+77}_{-140}$  \\
$e_c$                                   & 0                          & 0                             & 0 \\
$\omega_c$ (\degree)                    & 0                          & 0                             & 0\\
\enddata
\end{deluxetable*}

\section{Atmosphere}
\label{sec:atm}

We used our Bayesian Atmospheric Radiative Transfer code (BART,
\citealp{HarringtonEtal2021psjBART1, CubillosEtal2021psjBART2,
  BlecicEtal2021psjBART3}) to retrieve the atmosphere of
WASP-34b. BART consists of three main packages: Transit
\citep{RojoPhD}, a radiative transfer code that produces spectra from
a parameterized atmosphere model; Thermochemical Equilibrium
Abundances (TEA, \citealp{BlecicEtal2016apjsTEA}), which calculates
species abundances at each pressure and temperature in a planet's
atmosphere based on equilibrium chemistry; and MC$^3$
\citep{CubillosEtal2017ajRedNoise}, an MCMC routine wrapper. BART ties
these packages together to retrieve thermal profiles and abundances of
atmospheric constituents from eclipse or transit observations.

BART parameterizes the planetary thermal structure with the thermal
profile from \cite{LineEtal2013apjRetrievalI}. This model has five
free parameters: $\kappa$, the infrared Planck mean opacity;
$\gamma_1$ and $\gamma_2$, the ratios of Planck mean opacities in the
two visible streams to the infrared stream; $\alpha$, which splits
flux between the two visible streams; and $\beta$, which covers
albedo, emissivity, and heat redistribution. We also fit logarithmic
scale factors on the abundances of H$_2$O, CH$_4$, CO, and CO$_2$. All
parameter ranges are wide, and priors are uniform. Given the low
signal-to-noise of our data and the limited spectral coverage, we use
uniform abundances with respect to pressure. We include opacity from
the four aforementioned molecules \citep{RothmanEtal2010jqsrtHITEMP,
  LiEtal2015apjsCOLineList, HargreavesEtal2020apjsHITEMPch4} as well
as H$_2$ - H$_2$ collision-induced absorption.

Our spectrum is only two broadband photometric filters, so models are
prone to overfitting. We try several statistically- and
physically-motivated cases to determine what information we can learn
from our data:

\begin{enumerate}
\item All parameters free ($\kappa$, $\gamma_1$, $\gamma_2$, $\alpha$,
  $\beta$, and logarithmic scale factors for H$_2$O, CH$_4$, CO, and
  CO$_2$ abundances).
\item Since methane and CO$_2$ are not expected to be abundant at the
  equilibrium temperature of WASP-34b, we fix their abundances to
  6.93\ttt{-6} and 1.66\ttt{-7}, respectively. These are TEA-computed
  values at 0.1 bars pressure and the planetary equilibrium
  temperature of 1158 K, assuming 0 albedo and uniform heat
  redistribution. Thermal profile parameters and the other molecular
  abundances are left free to vary.
\item Same as 2,\ but the the CO mixing ratio is fixed to 4.53\ttt{-4}
  (thermochemical equilibrium as in case 2), since only the 4.5
  \microns\ filter is sensitive to CO abundance.
\item Same as 3,\ but the H$_2$O mixing ratio is fixed to 3.84\ttt{-6}
  (thermochemical equilibrium as in case 2).  Only the thermal profile
  parameters are free to vary.
\item Same as 4,\ but $\alpha = 0.0$ and $\gamma_2 = 1$, removing one
  visible stream.
\item Same as 5,\ but $\beta = 1$. This sets the irradiation
  temperature equal to the planet's equilibrium temperature, assuming
  zero albedo and perfect heat redistribution.
\item An isothermal atmosphere, where planetary temperature is the
  only free parameter.
\end{enumerate}

\noindent
Case 1 represents the most flexible model, cases 2 -- 4 make
simplifying assumptions about the atmospheric composition, and cases 5
-- 7 represent a range from complex to simple thermal profiles, all
with vertically-uniform molecular abundances. All cases include the
same opacity sources. As with the ``ramp'' in the light-curve
modeling, we use the BIC to determine which model is warranted by our
data (Equation \ref{eqn:bic}, Table \ref{tbl:bartbic}). We fit to both
the PLD and BLISS eclipse depths, separately, to compare results,
using the joint fits (Section \ref{sec:joint}), as the shared
parameters should lead to more accurate uncertainties.

\begin{deluxetable}{lrrrr}
\tablecaption{\label{tbl:bartbic} 
Atmospheric Fit BICs}
\tablehead{     & \multicolumn{2}{c}{BLISS} & \multicolumn{2}{c}{PLD}\\
\colhead{Case} & \colhead{BIC} & \colhead{$P_{21}$} & \colhead{BIC}  & \colhead{$P_{21}$}} 
\startdata
1    & 6.2384    &  0.0635       & 6.2387   & 0.0795  \\
2    & 4.8523    &  0.1270       & 4.8542   & 0.1589  \\
3    & 4.1590    &  0.1796       & 4.1591   & 0.2249  \\
4    & 3.4661    &  0.2540       & 3.5001   & 0.3127  \\
5    & 2.0800    &  0.5079       & 2.2010   & 0.5988  \\
6    & 1.3878    &  0.7180       & 1.5291   & 0.8379  \\
7    & 0.7251    &  ---          & 1.1753   & ---     \\
\enddata
\end{deluxetable}

\begin{figure*}
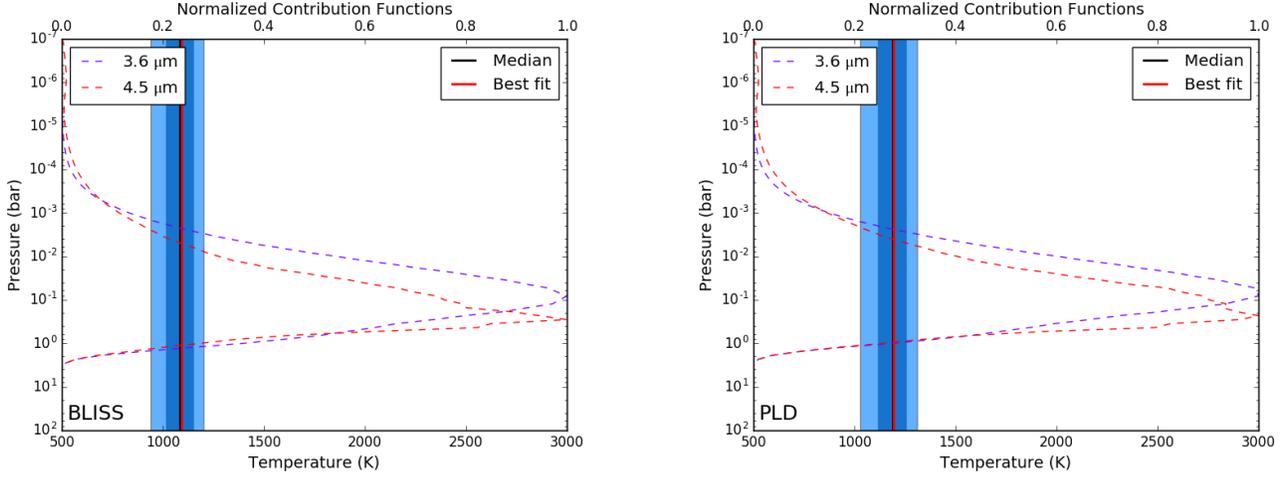

    \if\submitms y
        \setcounter{fignum}{\value{figure}}
        \addtocounter{fignum}{1}
        \newcommand\fignam{f\arabic{fignum}a.pdf}
    \else
        \newcommand\fignam{figs/bliss-pt.pdf}
    \fi
    \includegraphics[width=0.48\textwidth, clip]{\fignam}    
    \if\submitms y
        \setcounter{fignum}{\value{figure}}
        \addtocounter{fignum}{1}    
        \renewcommand\fignam{f\arabic{fignum}b.pdf}
    \else
        \renewcommand\fignam{figs/pld-pt.pdf}
    \fi
    \includegraphics[width=0.48\textwidth, clip]{\fignam}
    
    \caption{Lowest BIC BART-retrieved temperature-pressure
      profiles. Dark blue and light blue regions denote the 1 and
      2$\sigma$ boundaries, respectively. We have overplotted
      contribution functions for the two \textit{Spitzer} data points,
      which show the portion of the atmosphere probed by our
      retrieval. \textbf{Left:} The isothermal (case 7) profile
      retrieved from the BLISS eclipse depths. \textbf{Right:} The
      thermal profile retrieved using PLD eclipse depths (case 6).}
    \label{fig:pt}
\end{figure*}

The retrievals using the PLD and BLISS eclipse depths are very
similar. Cases 1, 2, and 3 result in fits with unconstained abundances
for all fitted molecules, with flat MCMC posteriors, indicating that
for any abundance within reasonable parameter bounds, there exists a
parameter set that fits equally well. The flat posteriors and a BIC
comparison show we are statistically justified in fixing the molecular
abundances to thermochemical equilibrium (Table
\ref{tbl:bartbic}). Likewise, our data are unable to support a
temperature structure as complex as case 4, with an uninformative
posterior distribution for $\alpha$. Cases 5, 6, and 7 have
informative (non-flat) marginalized posterior distributions for their
parameters, although cases 5 and 6 still overfit the data. With both
the BLISS and PLD eclipse depths, we find we are only justified in
fitting an isothermal atmosphere. We determine an isothermal
temperature of 1093 $\pm$ 66 K with BLISS (Figure \ref{fig:pt}, left)
and 1194 $\pm$ 66 K with PLD (Figure \ref{fig:pt}, right).

We tested MCMC convergence by ensuring that the Gelman \& Rubin test
was within 1\% of unity for all free parameters
\citep{GelmanRubin1992stascGRTest}. Also, we computed the Steps Per
Effectively Independent Sample and Effective Sample Size (SPEIS, ESS,
\citealp{HarringtonEtal2021psjBART1}) to verify that the posterior
distribution is well explored. Using the BLISS observations, we find a
SPEIS of 10, an ESS of 400, and a $68.3 \pm 2.3$\% ($1\sigma$)
credible region of [1026, 1159] K for the isothermal planet
temperature. For the PLD observations, we find a SPEIS of 10, an ESS
of 500, and a $68.3 \pm 2.1$\% credible region of [1130, 1263] K for
the isothermal planet temperature.

Per the BART license, the code version, inputs, outputs, and
output-processing scripts for these best-fitting atmospheres, with
step-by-step instructions to reproduce the results presented in this
work, are contained in a reproducible-research compendium which can be
found on Zenodo doi:10.5281/zenodo.5096510. The compendium also
includes the light-curve data, models, and diagnostic plots.

\begin{figure*}
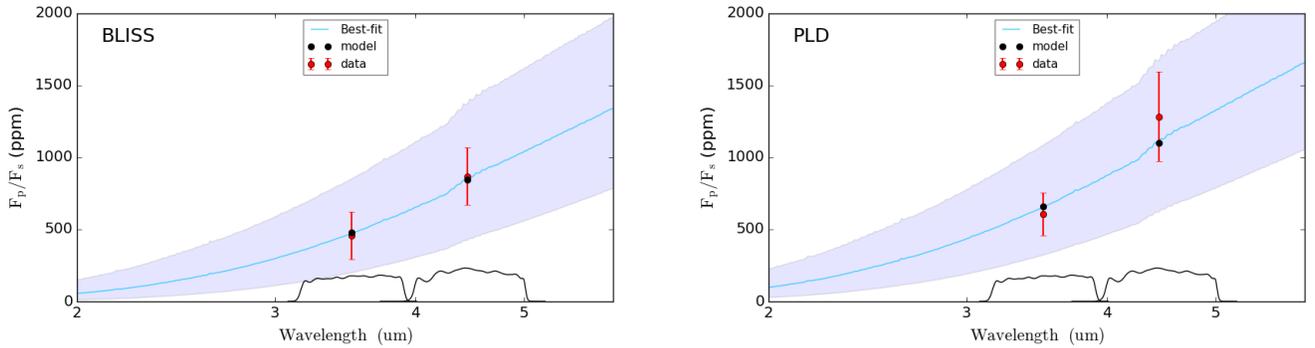

    \if\submitms y
        \setcounter{fignum}{\value{figure}}
        \addtocounter{fignum}{1}
        \newcommand\fignam{f\arabic{fignum}a.pdf}
    \else
        \newcommand\fignam{figs/bliss-spec.pdf}
    \fi
    \includegraphics[width=0.48\textwidth, clip]{\fignam}   
    \if\submitms y
        \setcounter{fignum}{\value{figure}}
        \addtocounter{fignum}{1}
        \renewcommand\fignam{f\arabic{fignum}b.pdf}
    \else
        \renewcommand\fignam{figs/pld-spec.pdf}
    \fi
    \includegraphics[width=0.48\textwidth, clip]{\fignam}
    
    \caption{Lowest BIC BART-retrieved spectra as planet-to-star flux
      ratio. The black dots are the spectrum integrated over the
      filters. The black dots, not the light blue line, should match
      the red data. The shaded region denotes the 3$\sigma$
      uncertainty on the best-fit isothermal atmosphere. The IRAC
      response functions are plotted along the $y$ axis (arbitrary
      units).  \textbf{Left:} Isothermal (blackbody) spectrum
      retrieved using the BLISS eclipse depths. Features are caused by
      the stellar spectrum (ATLAS9,
      \citealp{CastelliKurucz2004arxivStellarModels}); the planetary
      spectrum is a smooth blackbody curve. \textbf{Right:} The
      isothermal spectrum retrieved using the PLD eclispe depths.}
    \label{fig:spec}
\end{figure*}

\section{Discussion}
\label{sec:discussion}
With the number of Jupiters with measured emission increasing, many
studies have taken a statistical approach to exoplanet atmospheres,
both in transmission \citep[e.g.,][]{BaxterEtal2021aapTransitTrends}
and emission \citep[e.g.,][]{GarhartEtal2020ajHotJupiterStatChar,
  WallackEtal2021arxivSpitzerTrends}. Here we compare WASP-34b against
the literature of comparative exoplanetology to study how it fits into
observed trends.

\cite{GarhartEtal2020ajHotJupiterStatChar} noted a trend with slope
0.00043 $\pm$ 0.000072 in eclipse phase shift from a circular orbit
vs.\ orbital period. At WASP-34b's orbital period, we would expect a
shift of 0.00186 $\pm$ 0.00031 orbital phase. From joint light-curve
fits, we determined the eclipse phase shift to be 0.0018 $\pm$ 0.0007
(BLISS) and 0.0012 $\pm$ 0.0006 (PLD). The circularization of WASP-34b
agrees well with this trend.

Several studies have looked into trends in the ratio of 4.5 to 3.6
\microns\ brightness temperatures
\citep{KammerEtal2015apjCoolJupsTrends,
  WallackEtal2019ajCoolJupsTrends,
  GarhartEtal2020ajHotJupiterStatChar,
  WallackEtal2021arxivSpitzerTrends}. From our joint light-curve fits
we measured brightness temperature ratios of 1.05 $\pm$ 0.14 (BLISS)
and 1.13 $\pm$ 0.12 (PLD). WASP-34b may have a larger brightness
temperature ratio than other planets with a similar equilibrium
temperature, which generally fall below 1.0 (e.g., WASP-6b, WASP-8b,
WASP-39b, TrES-1b, \citealp{WallackEtal2021arxivSpitzerTrends}),
although the weak eclipses lead to large uncertainties. WASP-34b may
exhibit different chemistry than other warm Jupiters at the pressures
probed by these observations. For instance, 4.5 \microns\ CO emission
or 3.6 \microns\ CH$_4$ absorption could cause a redder slope. The
planet's unusual color is not attributable to its surface gravity
(log($g$) = 3.0) or host star metalliticy ([Fe/H] = -0.02 $\pm$ 0.10,
\citealp{SmalleyEtal2011aapWASP34b}), as these values are similar to
other warm Jupiters observed with \textit{Spitzer}.
  
\section{Conclusions}
\label{sec:conclusions}

We analyzed two \textit{Spitzer} observations of the exoplanet
WASP-34b using two light-curve modeling methods, BLISS and PLD, and
applying a modified eclipse model to account for the planet's high
impact parameter, demonstrating observational feasibility for
low-signal, grazing eclipses. The resulting eclipse depths, from joint
fits to both light curves, agree at $\leq1.1\sigma$ and eclipse
midpoint agrees at $0.7\sigma$ between the two methods. By minimizing
a combination of white and correlated noise, BLISS selects a fixed
photometry aperture radius but PLD prefers a variable aperture
radius. If the two models are forced to use the same light curve, the
resulting eclipse depths more closely match.

The measured eclipse midpoints further constrained the orbit of the
planet. We determined an eccentricity consistent with zero (0.0),
similar to previous works \citep{KnutsonEtal2014apjRVsearch,
  BonomoEtal2017aandaGAPS}. While $e \textrm{cos} \omega$ differs by
$1.7\sigma$ between fits to the BLISS and PLD eclipses, all other
fitted and derived orbital parameters are consistent between the two
orbital fits.

We also performed atmospheric retrieval on our measured eclipse
depths, separately for each light-curve modeling technique, using a
series of physically-motivated cases to determine what we could learn
from the data. For both BLISS and PLD, despite differences in the
eclipse depths, we preferred atmopsheric models that fixed molecular
abundances to thermochecmical equilibrium over those that fit the
abundances. Thus, we cannot constrain atmospheric constituents. We
find the best model, by BIC comparison, is an isothermal atmosphere at
$\sim$1100 -- 1200 K.

WASP-34b is somewhat unusual, with its density among the lowest 0.8\%
of planets with a measured radius and mass. The planet is redder than
other Jupiters with this equilibrium temperature, possibly indicating
unique chemistry, and the large scale height implied by its low
density makes it an attractive target for transit
studies. Unfortunately, the planet's grazing nature makes it difficult
to observe and characterize. Further improvement over the atmospheric
results presented here may be posssible with the \textit{Hubble Space
  Telescope}, at least in transit geometry, but additional eclipses to
constrain the dayside atmosphere and orbit likely must wait for the
\textit{James Webb Space Telescope}.

\section{Acknowledgements}

We thank the referees for their insightful comments and the resulting
improvements to this manscript.  We thank contributors to SciPy,
Matplotlib, and the Python Programming Language, the free and
open-source community, the NASA Astrophysics Data System, and the JPL
Solar System Dynamics group for software and services.  This work is
based on observations made with the {\em Spitzer Space Telescope},
which was operated by the Jet Propulsion Laboratory, California
Institute of Technology under a contract with NASA.  This work was
supported by NASA Planetary Atmospheres grant NNX12AI69G and NASA
Astrophysics Data Analysis Program grant NNX13AF38G. Jasmina Blecic is
supported by NASA through the NASA ROSES-2016/Exoplanets Research
Program, grant NNX17AC03G.

\facility{Spitzer (IRAC)}

\software{NumPy \citep{HarrisEtal2020natNumPy}, Matplotlib
  \citep{Hunter2007cseMatplotlib}, SciPy
  \citep{VirtanenEtal2020natmSciPy}, MC$^3$
  \citep{CubillosEtal2017ajRedNoise}, POET
  \citep[e.g.,][]{ChallenerEtal2021psjSystematics}, BART
  \citep{HarringtonEtal2021psjBART1, CubillosEtal2021psjBART2,
    BlecicEtal2021psjBART3}, RadVel \citep{FultonEtal2018paspRadVel}}

\bibliography{wasp34.bib}

\end{document}